\documentclass[
reprint, twocolumn, secnumarabic%
,tightenlines%
,amssymb,
amsmath,
nobibnotes,
nofootinbib, aps, prl, showpacs,showkeys]{revtex4}

\usepackage{subfigure}
\usepackage{tikz}
\usepackage{docs}%
\usepackage{bm}%
\usepackage{graphicx} %
\usepackage{rotating} %
\usepackage{hyperref} %
\usepackage{url} %
\usepackage{srcltx} %
\usepackage{float}
\expandafter \ifx \csname package@font\endcsname\relax\else
\expandafter \expandafter \expandafter
\usepackage
\expandafter \expandafter
\expandafter{\csname package@font\endcsname}%
\fi
\usepackage{epstopdf}
\usepackage{slashed}
\usepackage{ulem}
\usepackage{srcltx}
\usepackage{soul}
\usepackage{amsmath}
\usepackage{natbib} 
\usepackage{ulem} 

\usepackage{amsmath}
\usepackage{forest}

\usepackage{wasysym}



\usepackage{epstopdf}

\usepackage{enumerate}

\begin{document}

\title{Neutral current neutrino and antineutrino scattering off the polarized nucleon}

\author{Krzysztof M. Graczyk}
\email{krzysztof.graczyk@uwr.edu.pl}

\author{Beata E. Kowal}
	
\email{beata.kowal@uwr.edu.pl}

\affiliation{Institute of Theoretical Physics, University of Wroc\l aw, plac Maxa Borna 9,
50-204, Wroc\l aw, Poland}

\date{\today}%

\begin{abstract}

The elastic and inelastic neutral current $\nu$ ($\overline{\nu}$) scattering off the polarized nucleon is discussed. The inelastic scattering concerns the single-pion production process. We show that the spin asymmetries' measurement can help to distinguish between neutrino and antineutrino neutral current scattering processes. The spin asymmetries also encode information about a type of target. Eventually, detailed studies of the inelastic spin asymmetries can improve understanding of the resonant-nonresonant pion production mechanism.

\end{abstract}

\maketitle

\section{Introduction}

For the last few decades, considerable effort has been made to uncover the fundamental properties of neutrinos. One of the crucial tasks is to measure, with high accuracy, the neutrino oscillation parameters and the CP (charge conjugation and parity reversal) violation phase in the lepton sector. Indeed, it is one of the goals of ongoing experiments such as T2K \cite{sanchez_federico_2018_1295707}, or No$\nu$a \cite{sanchez_mayly_2018_1286758}. The measurement of the CP violation phase is essential not only for studies of neutrino and antineutrino properties but also can help understand the observed matter-antimatter asymmetry in the Universe \cite{Sakharov:1967dj,Fukugita:1986hr}. 

The CP violation phase has recently been determined by the T2K experiment~\cite{T2K:2019bcf} — this measurement is based on analyzing the neutrino and antineutrino oscillation data. A non-zero value of the CP phase means that neutrinos oscillate differently than antineutrinos. To study the oscillation phenomenon, one has to detect the interactions of neutrinos and antineutrinos with the nucleons and nuclei in the detectors, and one must be able to distinguish between neutrino and antineutrino processes to classify measured scattering events correctly.
Distinguishing between neutrino- and antineutrino-induced processes is also essential in detecting supernova neutrinos and antineutrinos \cite{Jachowicz:2004we}. Information about the energy spectrum of neutrinos and antineutrinos emitted during a supernova explosion is crucial for the development of the supernova theory \cite{Buras:2003sn}.

A neutrino is a neutral particle that interacts very weakly with matter. Therefore, measuring its interactions with nucleons or nuclei is a challenging task. We distinguish two types of neutrino interactions: charged current (CC) and neutral current (NC). In the first, the charged lepton is one of the products of interaction; in the other, there is no charged lepton in the final state. Both types of processes are detected in the long baseline neutrino oscillation experiments. In the case of supernova neutrinos (antineutrinos), only NC events are observed because neutrinos (antineutrinos) are of low energies, $10$ to $20$ MeV.

The history of studies of neutrino properties is inseparably connected with the investigation of fundamental interactions. For instance, discovering the NC interactions was essential for confirming the Glashow-Salam-Weinberg model for electroweak interactions. The first measurements of the NC neutrino and antineutrino scattering off nucleons and electrons were conducted by the Gargamelle experiment \cite{GargamelleNeutrino:1973jyy}. The observation of NC interactions resulted in the measurement of the Weinberg angle and the ratio of the nucleon $F_2$ structured functions obtained from electron and neutrino deep inelastic scattering off the nucleon. Certainly, the NC neutrino-matter interactions studies shall further discover the fundamental properties of weak interactions and matter.

It is trivial to distinguish between neutrino- and antineutrino-induced CC reactions when the charged lepton is in the final state. The lack of charged lepton in the final state in the NC neutrino-nucleon scattering makes detecting such events complicated. In this case, the event analysis is made based on the measurement of the recoil nucleon and other final hadronic particles. However, verifying if the measured nucleon is a product of neutrino or antineutrino processes is challenging. Another problem is distinguishing between elastic (El) and inelastic types of processes. The gross contribution to the inelastic scattering is from the processes in which a single pion is in the final state. However, for some events, the produced pion is either not visible in the detector or absorbed by the nuclear matter. Such measurements can be misidentified as the El contribution.

This paper focuses on NC neutrino and antineutrino scattering on a polarized target in the energy range characteristic for long baseline neutrino oscillation experiments such as T2K,  but we also consider supernova neutrino energies. Hence,  $E$ varies from about 10 MeV to several GeVs. In such energy range, there are two dominant types of processes for NC $\nu (\overline{\nu})$-$\mathcal{N}$ scattering ($\mathcal{N}$ denotes  proton or neutron), namely, elastic and single-pion production (SPP), in the last one there are nucleon and one pion in the final state.

In $\nu \mathcal{N}$ interactions, the polarization effects have been discussed for several decades \cite{SajjadAthar:2022pjt}  but mainly for the CC neutrino-nucleon/nucleus scattering. Polarization observables contain complementary, to the spin-averaged cross-section, information about the nucleon and nucleus structure~\cite{Akhiezer:1968ek,Donnelly:1985ry,Fasano:1992es,Nakayama:2019cys}. The first discussion of polarization properties in neutrino-induced processes appeared in the sixties and seventies \cite{Adler1963, Pais:1971er,LlewellynSmith:1971uhs} of the XX century. In 1965, Block \cite{Block:1965zol} announced one of the first experimental proposals to measure the polarization of the recoil nucleon in neutrino-deuteron scattering. Later, The polarization observables have been considered for the CC QE (quasielastic), CC SPP and deep inelastic  $\nu \mathcal{N}$ scattering \cite{Hagiwara:2003di,Kuzmin:2004yb,Kuzmin:2004ke,Graczyk:2004vg,Bourrely:2004iy,Graczyk:2019blt,Graczyk:2019opm,Graczyk:2019xwg,Graczyk:2021oyl} as well as for CC QE $\nu_\tau$-nucleus scattering~\cite{Graczyk:2004uy,Valverde:2006yi,Sobczyk:2019urm}. In the papers \cite{Bilenky:2013fra,Fatima:2018tzs,Graczyk:2019xwg,Tomalak:2020zlv,Graczyk:2021oyl} the sensitivity of the polarization asymmetries on the axial and strange nucleon form factors have been discussed. Recently,  Zaidi \textit{et al.} studied the impact of the nuclear effects on the tau polarization produced in CC deep inelastic $\nu_\tau/\overline{\nu}_\tau$-nucleus scattering \cite{Zaidi:2023hdd}. 

All papers cited above concern the CC interactions. The polarization effects in NC neutrino-nucleus scattering were studied by Jachowicz et al. \cite{Jachowicz:2004we}. It was shown that the measurement of recoiled nucleon polarization can help distinguish between neutrino and antineutrino interaction processes. These studies were extended in Refs. \cite{Jachowicz:2005np,Meucci2008}. Eventually,  Bilenky and Christova \cite{Bilenky:2013iua} pointed out that the polarization of the recoil nucleon in neutral current elastic (NCEl) interactions is sensitive to the axial form factor of the nucleon. Hence its measurement can provide information about the axial content of the nucleon.
\begin{figure}[t]
\centering{
\includegraphics[width=0.45\textwidth]{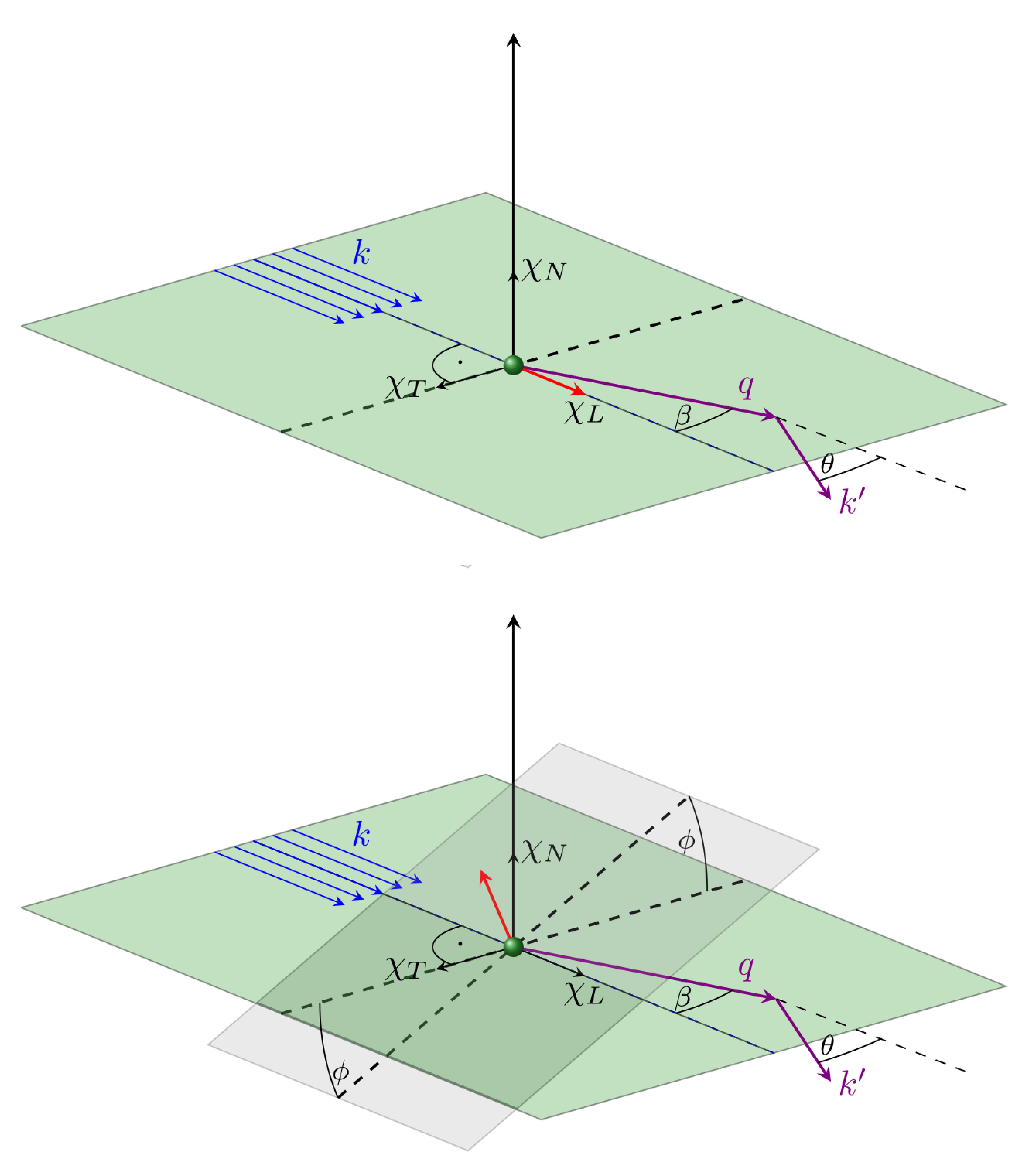}
\caption{In the top figure: neutrino scattering off the longitudinally (along the neutrino beam) polarized nucleon. In the bottom figure: the neutrino scattering off the nucleon polarized perpendicularly to the beam. The scattering plane has green color. The normal plane to the perpendicular polarization is gray. The red vector denotes the polarization of the nucleon. The polarization vectors $\chi_{L,T,N}$ are drawn in both panels.  
\label{Fig:Diagram}}}
\end{figure}

This paper continues our previous studies on polarization effects in the CC QE \cite{Graczyk:2019xwg} and SPP \cite{Graczyk:2017rti,Graczyk:2019blt,Graczyk:2021oyl} $\nu (\overline{\nu})$-nucleon scattering. We proposed a few types of spin asymmetry (SA) observables that contain nontrivial information about the nature of the interaction of neutrinos with nucleons. In Ref. \cite{Graczyk:2017rti}, we showed that the polarization of the outgoing nucleon, in the CC SPP $\nu \mathcal{N}$ scattering, hides the information about the relative phase between resonant-nonresonant background amplitudes. The following paper, Ref. \cite{Graczyk:2019blt}, revealed that target spin asymmetries are sensitive to the nonresonant background contribution. Eventually, for the CCQE $\nu_\mu \mathcal{N}$  scattering, we discussed three observables that had not been considered before, namely, target spin asymmetry, double and triple spin asymmetries \cite{Graczyk:2019xwg}. These observables turned out to be sensitive to the axial nucleon form factors, and their measurement can improve our knowledge about the axial contribution to the electroweak vertex of the nucleon. 

In the present paper, we shall show that for the neutrino (antineutrino) energies smaller than $1$~GeV, measurement of the target spin asymmetry for NC interactions allows one to distinguish between neutrino and antineutrino-induced processes. Moreover, the target SA brings information about a type of target nucleon that interacted with the initial neutrino. Eventually, the detailed analysis of the SAs can help to distinguish between El and SPP types of scattering events and bring information about El and SPP dynamics.

The paper is organized as follows: in Sec.~\ref{Sec:Formalism}, we discuss the necessary formalism and shortly review theoretical models for El and SPP interactions. In Sec.~\ref{Sec:Results}, we discuss numerical results. Sec.~\ref{Sec:Summary} summarizes the paper. We include three appendixes \ref{Appendix:El:Form-Factors}, \ref{Appendix:Asymetries:El} and \ref{Appendix:SPP} containing more details about El and SPP models.

\section{Formalism}
\label{Sec:Formalism}
\subsection{Spin asymmetries}

The present studies consider target spin asymmetry in NC El and SPP neutrino (antineutrino) -nucleon scattering processes. Namely, the neutrino or antineutrino collides with a polarized target,
\begin{equation}
\label{Eq:Target_process}
	 \nu(k) + \vec{\mathcal{N}} (p,s)  \to   \begin{cases}
  \nu(k') +  \mathcal{N}(p')   & \text{ El} \\
  \nu(k') +   \mathcal{N}'(p') + \pi(l) & \text{SPP}
\end{cases},
\end{equation}  
where $k^\mu= (E,\mathbf{k})$ and ${k'}^\mu = (E',\mathbf{k'})$ denote the four-momentum of the incoming and outgoing neutrinos (antineutrinos); $p$ and $p'$ are the four-momenta of the target nucleon, and outgoing nucleon, respectively; $l$ is the pion four-momentum; $s$ is the spin four-vector of the target nucleon. We work in the laboratory frame: $p^\mu = (M,0)$ ($M$ is nucleon averaged mass).

The differential cross section for (\ref{Eq:Target_process}) type of the process reads
\begin{equation}
d\sigma (s_{\mu} ) = d\sigma_0   \left(1  + \mathcal{T}^\mu s_{\mu} \right),
\end{equation} 
where $\mathcal{T}^\mu$ is the target spin asymmetry four-vector with three independent components; $d\sigma_0$ is half of the spin averaged cross-section. 

To compute the components of $\mathcal{T}^\mu$ we introduce the spin basis (see Fig.~\ref{Fig:Diagram}):
\begin{eqnarray} 
\label{basis_target_L}
\chi_L^\mu    & = & \frac{1}{E} \left(0,\textbf{k} \right), \\
\chi_T^\mu       & = &
\label{basis_target_T}
\left(0,\frac{\textbf{k}\times(\textbf{k}\times \textbf{q})}{|\textbf{k}\times(\textbf{k}\times \textbf{q})|}\right), \\
\label{basis_target_N}
\chi_N^\mu    & = & \left(0,\frac{\textbf{k}\times \textbf{q}}{|\textbf{k}\times \textbf{q}|}  \right), 
\end{eqnarray}
where $\mathbf{q} =\mathbf{k} - \mathbf{k'} $. The $\chi_L^\mu$ is the vector longitudinal along the neutrino beam; $\chi_N^\mu$ is normal to the scattering plane, and $\chi_T^\mu$, transverse component that lies in the scattering plane.

With the above choice of basis, there are three independent components of the target spin asymmetry, namely:
\begin{equation}
\mathcal{T}^X \equiv    \chi_X^\mu  \mathcal{T}_{\mu}, \quad X=L, T, N,
\end{equation}
and the target spin asymmetries are given by the ratio
\begin{equation}
\label{Eq:Pratio}
R(d \sigma,s_X; A, B,...)  
=\frac{\displaystyle \sum_{c=\pm1}  c\, {d\sigma}(A,B,...,c \,s_X)}{\displaystyle \sum_{c=\pm1}  {d\sigma}(A,B,...,c \, s_X   )},
\end{equation}
where $A$, $B$, ... stand for the kinematic variables the $d\sigma$ depends on, $s_X$ is the spin vector.

In this paper, we compute ratios of the total cross sections, which are given by
\begin{equation}
\label{Eq:Asymmetry-Single}
 \mathcal{T}_{X}(E) \equiv R\left(\sigma,\chi_X; E\right),
\end{equation} 
where $\sigma$ is the total cross section.
However, in the appendix \ref{Appendix:Asymetries:El}, we give the analytic formulas for 
$R\left(d\sigma/dt,\chi_X; E, t\right)$ ($t=q^2$) computed for NC El scattering, that can be used to compute (\ref{Eq:Asymmetry-Single}). The formulas for NC SPP asymmetries are too complicated and too long to present in the paper. Note that we used FORM language \cite{Vermaseren:2000nd} to obtain the analytic expressions for asymmetries. The numerical computations have been conducted in C++ language.

We consider two scenarios for the polarization of the nucleon target: along the neutrino beam and perpendicular to the neutrino beam.
Note that the direction of the neutrino beam in the long and short baseline experiments is fixed. Hence,  polarizing the beam along the neutrino beam is a natural option, and it does not introduce additional complications to the analysis. This scenario is shown in the top diagram of  Fig. \ref{Fig:Diagram}. In the second scenario, shown in the bottom diagram of Fig. \ref{Fig:Diagram}, we consider the polarization of the nucleon target perpendicular to the beam. However, the scattering plane spanned by lepton vectors defines the two spin vectors, normal ($\chi_\mu^N$) and transverse ($\chi_\mu^T$). Then, the linear combination of the normal and transverse components will give the measured perpendicular ($\mathcal{T}_\perp$) spin asymmetry. 

Note that the normal component for El scattering vanishes if one assumes that the nucleon's vector and axial form factors are real. The imaginary contribution to the form factors (on the tree level) can only be possible for the types of neutrino-nucleon interactions that go beyond the Standard Model. 

In contrast to El scattering, the normal component for the SPP processes does not vanish. However, we found that this contribution is of the order of $10^{-3}$ and would be difficult to measure.
Hence, the transverse component for both El and SPP processes fully determines the target spin asymmetry perpendicular to the neutrino beam, namely,
$\mathcal{T}^T \approx \sin \phi \mathcal{T}_\perp$, where $\phi$ is defined in Fig. \ref{Fig:Diagram}.
\begin{figure} 
\includegraphics[width=0.95\linewidth]{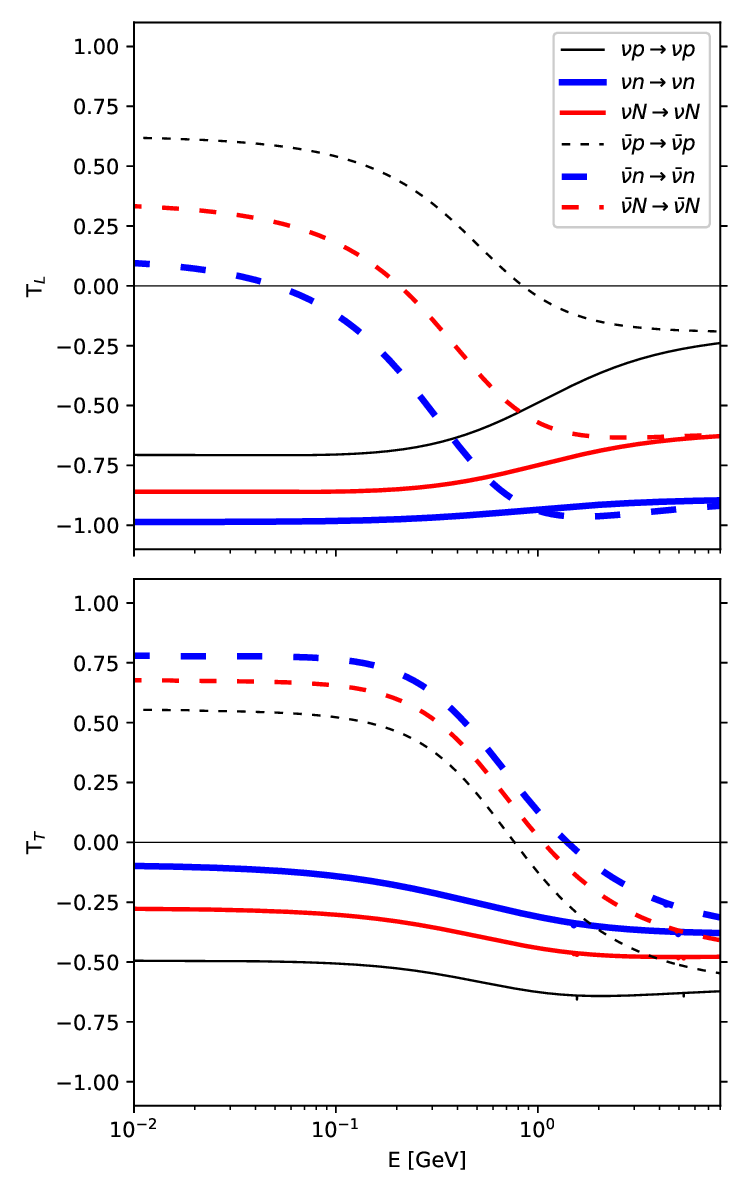}
\caption{Target spin asymmetry for NCEl. The solid/dashed line denotes the spin asymmetries computed for the neutrino/antineutrino -proton (black line), -neutron (blue line), and isoscalar target (red line) scattering. In the top/bottom panel  $\mathcal{T}^{L}(E)$/$\mathcal{T}^{T}(E)$ is plotted. \label{Fig:NC_EL_total}} 
\end{figure}

\subsection{Cross-section models}

According to the standard model, the NC  types of interactions are described by the density Lagrangian~\cite{Alberico:2001sd}:
\begin{equation}
\mathcal{L}_{NC}
=
- \frac{g}{2\cos\theta_W} \mathcal{J}_\alpha^{NC} Z^\alpha + h.c.,
\end{equation}
where $G_F/\sqrt{2} = g^2/8 m_W^2$, $G_F$ -- Fermi constant; $g$ weak coupling constant; $m_{W} = \cos \theta_W m_{Z}$ is mass of the $W^\pm$ and $m_Z$ is the mass of $Z^0$ boson, and $Z^\mu$ is the gauge field; $\theta_W$ is the Weinberg angle. 

 \begin{figure*}
   \includegraphics[width=0.9\linewidth]{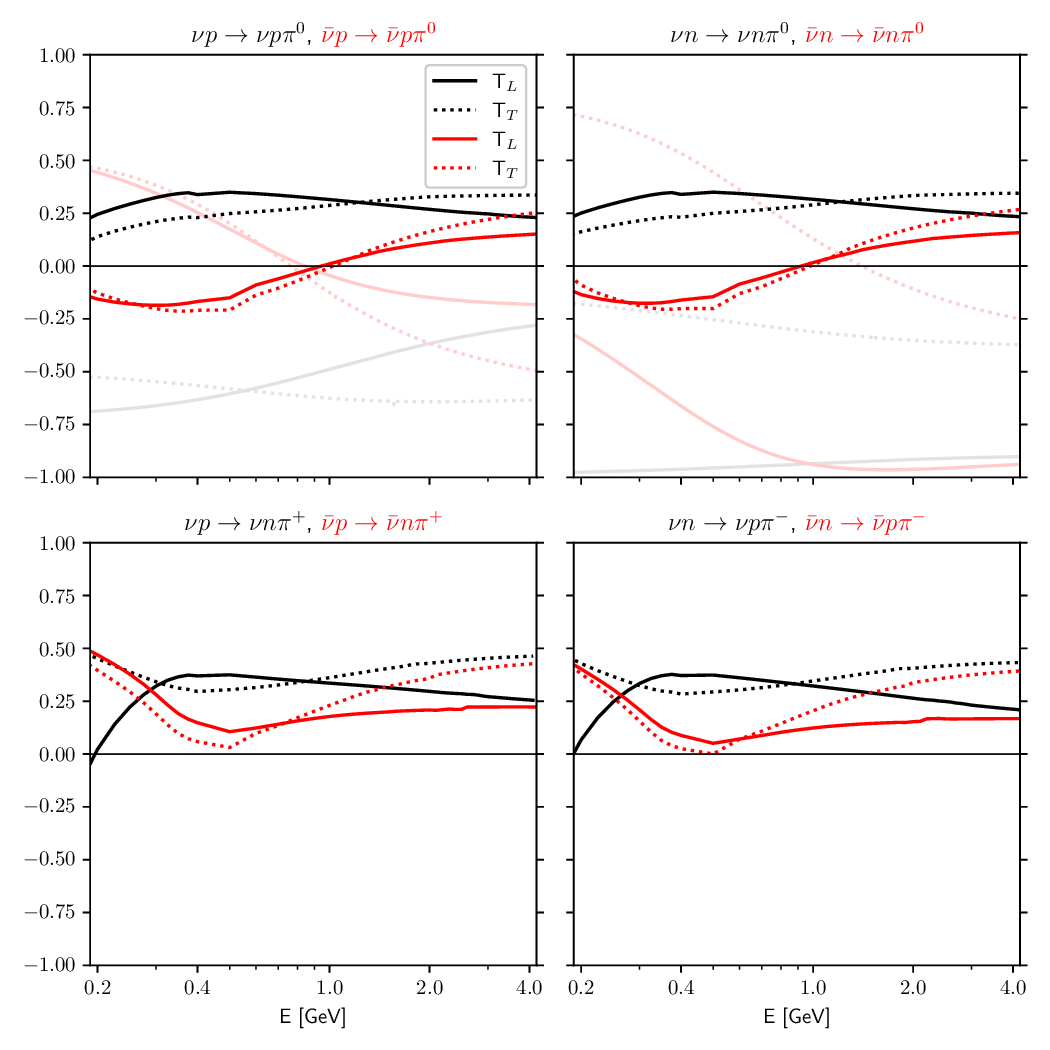}
  \caption{Target spin asymmetry for NC SPP (full model). The solid/dotted line corresponds to the $\mathcal{T}_{L/T}$ component of the spin asymmetries. The black/red line denotes the spin asymmetries computed for $\nu$/$\overline{\nu}$  scattering off the nucleon. In the top panels, we plot the corresponding NCEl target spin asymmetries in the background.
\label{Fig:SPP:SAs}} 
\end{figure*}
	
In a laboratory frame, the differential cross-section for NCEl  and NC SPP scattering processes reads
\begin{equation}
		\label{Eq:cross-section}
		{d\sigma}
		\sim  \mathrm{H}^{\mu\nu}_{NC}\mathrm{L}_{\mu\nu},
	\end{equation}
where $L_{\mu\nu}$ is the leptonic tensor that has the form		  
\begin{equation*}
\mathrm{L}_{\mu\nu} =    
8 \left(k^\nu {k'}^\mu+k^\mu {k'}^\nu -g^{\mu \nu} k\cdot k' \pm  i \epsilon^{\mu\nu \alpha\beta}k_\alpha {k'}_\beta \right).
\end{equation*}
Sign $\pm$ corresponds to neutrino/antineutrino scattering. The hadronic tensor has the form 
\begin{equation}
	\mathrm{H}_{NC}^{\mu\nu} =
	J_{NC}^\mu  {J_{NC}^{\nu}}^{*} , 
\end{equation}
where $J_{NC}$ is the expectation value of the hadronic current $\mathcal{J}_{NC}$.
\begin{figure*}
	   \includegraphics[width=0.9\linewidth]{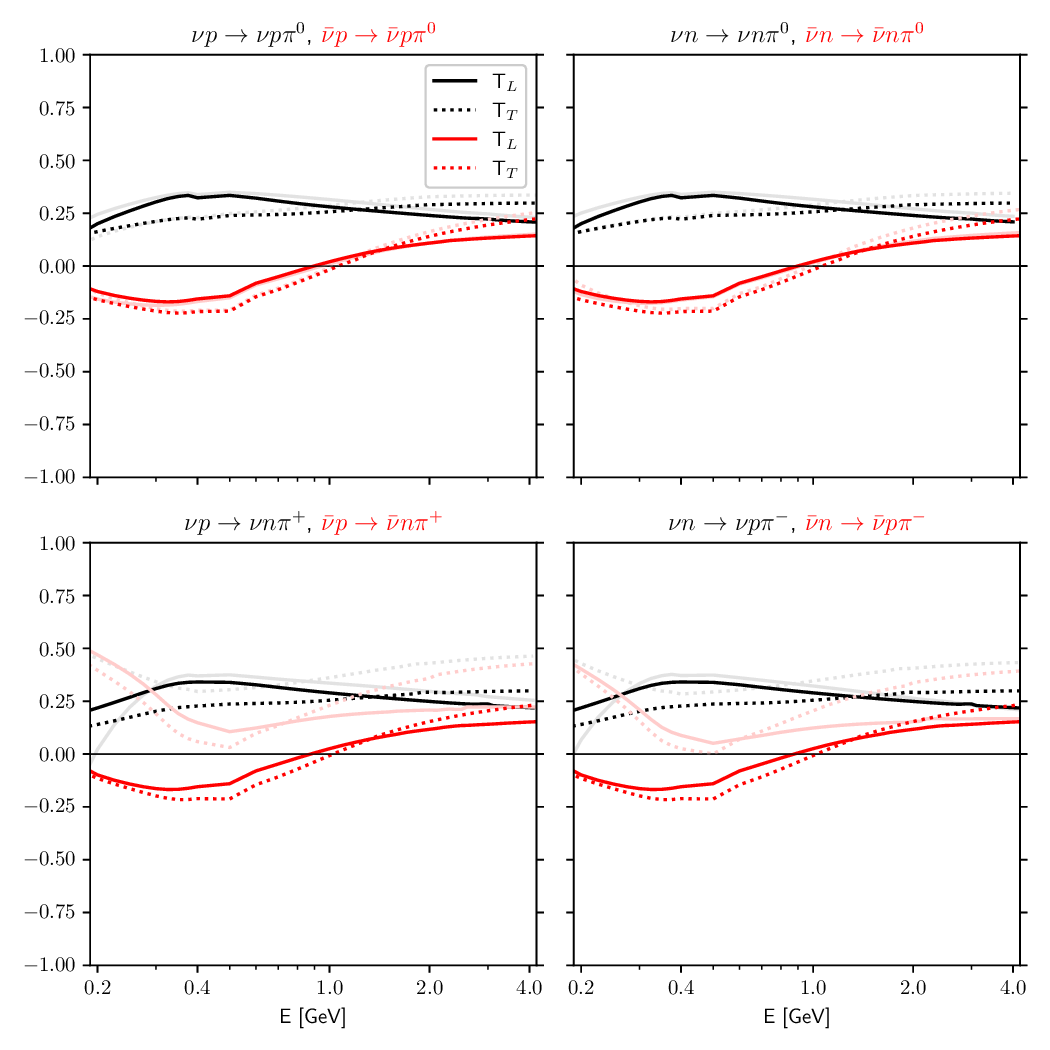}
  \caption{Target spin asymmetry for NC SPP but only $\mathcal{N}\to \Delta$ contribution described by DP and CDP diagrams. The solid/dotted line corresponds to the $\mathcal{T}_{L/T}$ component of the spin asymmetries. The black/red line denotes the spin asymmetries computed for $\nu$/$\overline{\nu}$  scattering off the nucleon. In the background, the corresponding full SPP model spin asymmetries are shown.
  \label{Fig:SPP:SAs_Delta}} 
\end{figure*}

To compute the cross-section, we need to construct the hadronic currents for both types of interaction. Derivation of the hadronic tensor for NCEl $\nu \mathcal{N}$ scattering is similar to the CCQE (see Sec. II of Ref. \cite{Graczyk:2019xwg}). The main difference lies in the parametrization of the form factors and kinematics. We provide some details in Appendix~\ref{Appendix:El:Form-Factors}. Note that we show the numerical results for six types of processes: $\nu \mathcal{N} \to \nu \mathcal{N}$ and  $\overline \nu \mathcal{N} \to \overline \nu \mathcal{N}$ scatterings, where $\mathcal{N} =$ proton (p), neutron (n), as well as for $\nu N \to \nu N$ and $\overline \nu N \to \overline \nu N$, where $N$ denotes the isoscalar target. To compute the cross-section for neutrino scattering off the isoscalar target, we assume that the target contains the same number of protons and neutrons and the cross-section reads $d\sigma^N = (d \sigma^p + d\sigma^{n})/2$.

We adapt the model from Hernandez \textit{et al.} \cite{Hernandez:2007qq} to compute the NC SPP cross-section. The model describes the neutrino-deuteron scattering data well, and its vector contribution can be fitted to the electroproduction data \cite{Graczyk:2014dpa}. 
The sum of seven amplitudes gives the total amplitude for the SPP induced by $\nu \mathcal{N}$ interaction. Two amplitudes, denoted by DP ($\Delta$ pole) and CDP (crossed $\Delta$ pole), contain a contribution from nucleon$\to \Delta(1232)$ (resonance) transition. The contributions from the nucleon excitation to heavier resonances are small in the energy range considered in the present studies. The remaining five amplitudes (NP - nucleon pole, CNP - crossed nucleon pole, PF - pion in flight, CT - contact term, and PP - pion pole)  describe the nonresonant background contribution. 

Similarly, as in the NCEl case, computing the NC SPP cross-section is very similar to those performed for CC SPP, see Sec. II and Sec. III of Ref. \cite{Graczyk:2017rti}. The main difference lies in describing the elementary vertices (form factors and Clebsch-Gordan coefficients) and kinematics. Some details are given in Appendix~\ref{Appendix:SPP}. 

There are four variants of the SPP neutrino-induced process: 
	\begin{eqnarray}
    \nu p   & \to &  \nu   p\pi^0, \\   
	\nu n  & \to &  \nu   n \pi^0 ,\\  
	\nu p  & \to &  \nu   n \pi^+, \\   
	\nu n  & \to & \nu   p \pi^-,
    \end{eqnarray}
    and corresponding four SPP antineutrino-induced process:
	\begin{eqnarray}
     \bar{\nu}  p  & \to &  \bar{\nu}   p  \pi^0, \\ 
	\bar{\nu}  n  & \to &  \bar{\nu}   n   \pi^0, \\ 
	\bar{\nu}  p  & \to &  \bar{\nu}   n   \pi^+ , \\  
	\bar{\nu}  n  & \to &  \bar{\nu}   p   \pi^- .
  \end{eqnarray}
In contrast to El scattering, various approaches have been developed to describe the SPP. They differ in the treatment of the resonance and nonresonant contributions, and there is a need for providing new observables that help to testify the models \cite{Graczyk:2017rti}.
\begin{figure*}
   \includegraphics[width=0.9\linewidth]{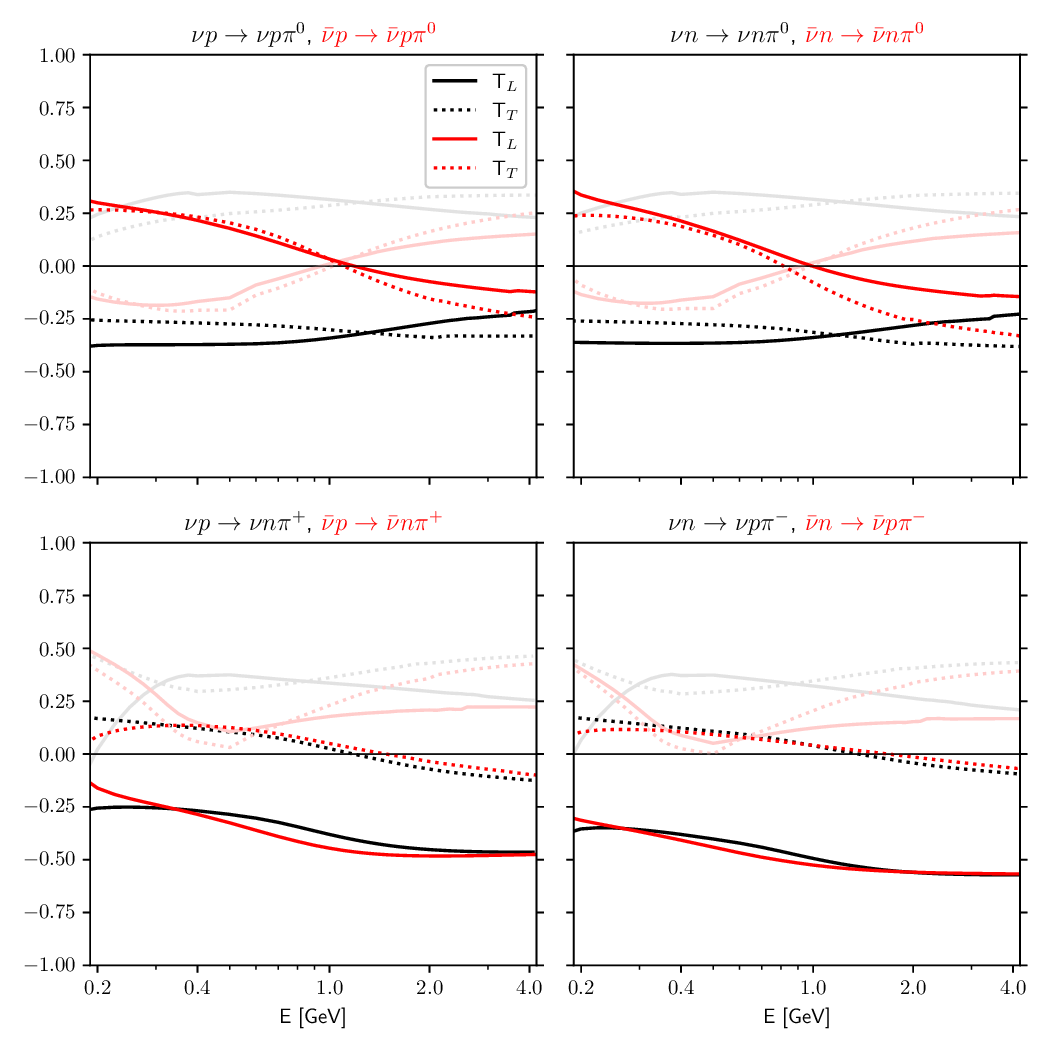}
  \caption{Target spin asymmetry for NC SPP but only nucleon-pole contribution described by NP and CNP diagrams. The solid/dotted line corresponds to the $\mathcal{T}_{L/T}$ component of the spin asymmetries. The black/red line denotes the spin asymmetries computed for $\nu$/$\overline{\nu}$  scattering off the nucleon. In the background, the corresponding full SPP model spin asymmetries are shown.
  \label{Fig:SPP:SAs_NP_CNP}} 
\end{figure*}

\section{Results}
\label{Sec:Results}

We begin the discussion of the results from the El scattering. In Fig.~\ref{Fig:NC_EL_total}, we plot the target spin asymmetries defined by the ratios of the total cross sections. We consider El scattering on neutron, proton, and isoscalar targets. The energies vary from $E = 0.01$ GeV to  $4$ GeV, which includes supernova and accelerator energy ranges of neutrinos. Notably, below $\nu$ ($\overline{\nu}$) energy approximately $E\sim0.7$ GeV, the transverse components of SAs for $\nu$ and $\overline{\nu}$ interactions differ in sign and energy dependence. Conversely, the longitudinal components of $\nu$ and $\overline{\nu}$ target SAs, computed for neutron, have the same sign in the entire range. Almost the same property holds for the longitudinal component computed for the isoscalar target. Indeed, in this case, the sign difference for neutrino/antineutrino is seen only at the low energy range. Eventually, the difference in the sign for $\nu$ and $\overline{\nu}$ asymmetries is exhibited for the $E<1$~GeV for the proton target. The disparities between the SAs for neutrinos and antineutrinos gradually vanish when beam energy increases. Moreover, for neutrino (antineutrino) energies $E\sim 5$~GeV, the asymmetries tend to converge to some fixed values specified for each target type.

In the analysis of the SPP, we distinguish $\pi^0$ and $\pi^\pm$ production processes. In Fig.~\ref{Fig:SPP:SAs}, we show the longitudinal and transverse components of the SA for both types of processes. In the case of $\pi^0$  production, the longitudinal and transverse components are of the same order and sign. Similarly, as in the NCEl case,  the $\nu$ and $\overline{\nu}$ target spin asymmetries (longitudinal and transverse) have opposite signs for $E<1$ ~GeV.  When the energy of the $\nu$ ($\overline{\nu}$) grows, the difference between SAs for neutrino and antineutrino disappears. In the $\pi^0$ production process, the final nucleon has the same isospin as the initial one. From that perspective, there is a similarity between   $\pi^0$ production processes and NCEl ones. Hence in the top panels of Fig.~\ref{Fig:SPP:SAs}, in the background, we plot the NCEl SAs. As can be noticed, the analysis of energy dependence and signs of SPP SAs allows one to distinguish between elastic and SPP types of process.
For the SPP processes in which the target changes the identity and the charged pion is the final state, in contrast to $\pi^0$ production, both components of asymmetries for neutrino and antineutrino scattering have the same sign (positive) and similar energy dependence.

Note that the dominant contribution to the target spin asymmetries for SPP comes from resonance $\mathcal{N}\to \Delta (1232)$ transition, which is illustrated in  Fig.~\ref{Fig:SPP:SAs_Delta}. However, the background terms visibly contribute to the SAs. Indeed, in Fig.~\ref{Fig:SPP:SAs_NP_CNP}, we show the SA's computed only for diagrams NP and CNP. These diagrams correspond to the process at which the elementary interaction between neutrino (antineutrino) is the same as in NCEl, but the nucleon emits the pion. In this case, for energies below $1$~GeV, the signs of the SA's for $\pi^0$ production processes and El are negative for neutrino and positive for antineutrino scattering processes. 

For the remaining two SPP processes ($\pi^\pm$ production), the sign of the SAs depends on the type of polarization component rather than the initial lepton type. Altogether shows that the target spin asymmetries in NC SPP interactions are sensitive to the amplitude content and seem to contain valuable information about the dynamical structure of neutrino-nucleon interaction. 

\section{Summary}
\label{Sec:Summary}

It has been demonstrated that the target spin asymmetries for neutral current neutrino-nucleon and antineutrino-nucleon interactions differ in sign and energy dependence. Indeed, at energies below $0.7$ GeV, the transverse and partially longitudinal SA components for El and SPP processes take different sign values for $\nu-$ and $\overline{\nu}$- induced processes. An analogous property reveals SA's transverse and longitudinal components for $\pi^0$ production. However, SPP SAs take opposite signs to their counterparts from El scattering. A detailed analysis of the energy dependence of the elastic SAs can provide information about the type of the initial target. Eventually, the SPP spin asymmetries also contain information about the resonance-nonresonant content of scattering amplitudes. Hence, their investigation should contribute to studies of the fundamental properties of the neutrino-nucleon interactions in the 1~GeV energy range.

We conclude that measuring $\nu$ and $\overline \nu$ scattering on polarized target can discriminate between neutral current neutrino- and antineutrino-induced processes\footnote{If the neutrino has a Majorana nature, then one discriminates between left-handed and right-handed neutrino-induced processes.}. This property can be beneficial for measuring supernova neutrinos with so low energy that they can only interact via neutral currents. Disparities between neutrino and antineutrino interaction processes are also crucial for determining the CP violation phase. Hence, a new data type should result in a better determination of the oscillation parameters. Eventually,  the polarization observables' analysis can help distinguish between elastic and SPP types of events and constrain the theoretical models that describe them.

\section*{Acknowledgments}

 \begin{acknowledgments}

This research was funded in whole or in part by National Science Centre (UMO-2021/41/B/ST2/02778).

Authors supported by National Science Centre (UMO-2021/41/B/ST2/02778).

K.M.G is partly supported by the program ”Excellence initiative—research university” (2020-2026 University of Wroclaw).

A part of the algebraic calculations presented in this paper has been performed using FORM language \cite{Vermaseren:2000nd}.

The calculations have been carried out in Wroclaw Centre for Networking and Supercomputing (\url{http://www.wcss.wroc.pl}), grant No. 268.
  \end{acknowledgments}

\appendix

\section{Form factors for elastic scattering}

\label{Appendix:El:Form-Factors}

The hadronic current has vector ($V$) - axial ($A$) structure
\begin{equation}
	    J_{NC;\mathcal{N}}^\mu  = \bar{u}_{\mathcal{N}} \left(\gamma_{\alpha}  \widetilde{F}^{\mathcal{N}}_1 + \frac{i}{2M} \sigma_{\alpha \beta} q^{\beta} \widetilde{F}^{\mathcal{N}}_2 + \gamma_{\alpha} \gamma_5 \widetilde{F}^{\mathcal{N}}_A  \right)u_{\mathcal{N}},
\end{equation}
where $\mathcal{N}=p,n$.
  
The form factors for nucleon have the following form 	\begin{eqnarray}
	\widetilde{F}^{p(n)}_{1,2}
	&=& + (-)  \left(1 -  2\sin^2\theta_W\right)  \frac{F^V_{1,2}}{2}   -  \sin^2\theta_W  F^S_{1,2} \nonumber \\ \\
	{F}^{V(S)}_{1,2} &=&      F^p_{1,2}  - (+) F^n_{1,2}.   \nonumber
	\end{eqnarray}
   $F_{1,2}^{p(n)}$ is proton (neutron) form factor, fit II from Ref.~\cite{Alberico:2008sz} (for the proton and neutron,  Eqs.~40, 47).

The axial form factor for proton (neutron) for NC reads	
\begin{equation}
\widetilde{F}^{p(n)}_{A}= +(-)\frac{1}{2}  F_{A} ,
\end{equation}	
where $F_A$ is CCQE axial form factor. We assume the dipole parametrization
\begin{equation}
F_A (t) = 1.2723 ( 1 - t/M_A^2)^{-2}, \quad M_A = 1~\mathrm{ GeV}.
\end{equation}

\section{Polarization asymmetries for El scattering}
\label{Appendix:Asymetries:El}

Here we give the spin asymmetry formulas 
for $R\left(d\sigma/dt,\chi_X; E, t\right)$. Note that to compute $R\left(\sigma,\chi_X; E\right)$, one computes ratio of the integrals over $t$ of the numerator and  denominator of $R\left(d\sigma/dt,\chi_X; E, t\right)$.

Let 
\begin{equation}
t = q^2 = (k-k')\cdot(k-k'),
\end{equation}
where $q^\mu = (\omega, \mathbf{q})$.

Longitudinal target asymmetry, $R\left(d\sigma/dt,\chi_L; E, t\right)$:
\begin{eqnarray}
\mathcal{T}_L&=& \frac{1}{4 E  \mathcal{I} } 
\left[ \right. 8 t x \tilde{F}_1^2 \left(4 E^2 M+t (E+M)\right)
-8 E t^2 x \tilde{F}_2^2 \nonumber\\
&+& 8 M t x \tilde{F}_1 \tilde{F}_2 \left(4 E^2+t\right)\nonumber\\
&-&8 M t \tilde{F}_2  \tilde{F}_A \left(t-4 E^2\right) \nonumber\\
&-&16 \tilde{F}_1  \tilde{F}_A \left(8 E^3 M^2+t^2 (E+M)+2 E M t (2 E+M)\right) \nonumber\\
&-&8 t x (-E-M) \tilde{F}_A^2 (4 E M+t) 
 \left. \right].  
 \label{Eq:T_L}
\end{eqnarray} 
Transverse target asymmetry, $R\left(d\sigma/dt,\chi_T; E, t\right)$:
\begin{eqnarray}
\mathcal{T}_T&=& \frac{\sin\beta |\mathbf{p'}| }{  \mathcal{I} }\left[\right. 
4 \tilde{F}_2 \tilde{F}_A \left(-4 E^2 M-E t+M t\right)\nonumber\\
&+& 8 M  \tilde{F}_1  \tilde{F}_A (2 E M+t)
+ \tilde{F}_1\tilde{F}_2 (4 E t x-4 M t x) \nonumber\\
&-&4 M x \tilde{F}_A^2 (4 E M+t)+4 E t x \tilde{F}_2^2-4 M t x \tilde{F}_1^2
\left. \right].
\label{Eq:T_T}
\end{eqnarray}
where $x=\pm$ for neutrino/antineutrino,
$$2E(\omega-|\mathbf{q}|\cos\beta)=2 k \cdot q =t-(k-q)^2=t=-2 M \omega$$
$\beta$ - is an angle between $\mathbf{k}$ and $\mathbf{q}$.

$\mathcal{I}$  is the contraction of the leptonic and hadronic tensors and it reads
\begin{eqnarray}
\mathcal{I} &=&2 \tilde{F}_1^2 \left(8 E^2 M^2+2 M t (2 E+M)+t^2\right)\nonumber\\
&+&t \tilde{F}_2^2 \left(t-\frac{2 E (2 E M+t)}{M}\right)\nonumber\\
&+&4 t^2  \tilde{F}_1 \tilde{F}_2  - 4 t x (  \tilde{F}_1 + \tilde{F}_2  )\tilde{F}_A (4 E M+t)\nonumber\\
&+&2 \tilde{F}_A^2 \left(8 E^2 M^2+4 E M t+t \left(t-2 M^2\right)\right). 
\end{eqnarray}

\section{Details of implementation of NC SPP}
\label{Appendix:SPP}

In the table below, we include, for each process, the weight with which a given diagram contributes to the total amplitude, the form factors for the $\mathcal{N}\to \Delta$ transition, and the nonresonant background terms. For direct comparison, we keep charged current (CC) terms.

\begin{widetext}
{\scriptsize

		\begin{equation*}
	\begin{array}{ |l||c|c||c|c|c|c|} 
 	\hline
 \mbox{} &	\mbox{CC }  &  	\mbox{CC } & 	\mbox{NC} & \mbox{NC } & \mbox{NC} & \mbox{NC} \\
	 &	\mbox{ $\nu p \rightarrow l^- p \pi^+$}  &  	\mbox{ $\nu n \rightarrow l^- n \pi^+$} & 	\mbox{ $\nu p \rightarrow \nu p \pi^0$} & \mbox{ $\nu n \rightarrow \nu n \pi^0$} & \mbox{ $\nu p \rightarrow \nu n \pi^+$} & \mbox{ $\nu n \rightarrow \nu p \pi^-$}\\
	\hline
	\mbox{NP} &	0 & 1 & \frac{1}{\sqrt{2}}  & \frac{1}{\sqrt{2}} & 1 & -1\\
	          &	F_{1,2}^V & F_{1,2}^V & \frac{ F_{1,2}^V }{2}(1-2 s_W^2 )  - s_W^2 F_{1,2}^S   & \frac{F_{1,2}^V }{2} (1-2 s_W^2 )  + s_W^2 F_{1,2}^S &  \frac{ F_{1,2}^V }{2}(1-2 s_W^2 )  - s_W^2 F_{1,2}^S   & \frac{ F_{1,2}^V }{2} (1-2 s_W^2 )  + s_W^2 F_{1,2}^S \\
	 	   &	F_{A} & F_{A} & \frac{1}{2} F_{A}    & \frac{1}{2} F_{A}     & \frac{1}{2} F_{A}     & \frac{1}{2} F_{A}   \\
	 \hline
	\mbox{CNP} & 1 & 0 & \frac{1}{\sqrt{2}}  & \frac{1}{\sqrt{2}} & -1 & 1\\
	          &	F_{1,2}^V & F_{1,2}^V & \frac{ F_{1,2}^V }{2}(1-2 s_W^2 )  - s_W^2 F_{1,2}^S   & \frac{F_{1,2}^V }{2} (1-2 s_W^2 )  + s_W^2 F_{1,2}^S &  \frac{ F_{1,2}^V }{2}(1-2 s_W^2 )  + s_W^2 F_{1,2}^S   & \frac{ F_{1,2}^V }{2} (1-2 s_W^2 )  - s_W^2 F_{1,2}^S \\
	 	   &	F_{A} & F_{A} & \frac{1}{2} F_{A}    & \frac{1}{2} F_{A}     & \frac{1}{2} F_{A}     & \frac{1}{2} F_{A}   \\
	 \hline 
 	\mbox{PF} &	1 & -1 & 0  & 0 & -2 & 2\\
 		          &		F_{1}^V & F_{1}^V & --  & -- & \frac{ F_{1}^V }{2} (1-2 s_W^2 )  & \frac{ F_{1}^V }{2} (1-2 s_W^2 ) \\
	 \hline
 	\mbox{CT} &	1 & -1 & 0  & 0 & -2 & 2\\
 		 &	F_{1}^V & F_{1}^V & --  & -- & \frac{ F_{1}^V }{2} (1-2 s_W^2 )  & \frac{ F_{1}^V }{2} (1-2 s_W^2 ) \\
 	       & F_{\rho} & F_{\rho}  & --  & -- & \frac{1}{2}  F_{\rho}  & \frac{1}{2}  F_{\rho} \\
	 \hline
	\mbox{PP} & 1 & -1 & 0  & 0 & -2 & 2\\
  &   F_{\rho}   &   F_{\rho}   & --  & -- & \frac{1}{2}  F_{\rho}  & \frac{1}{2}  F_{\rho}  \\
		  & & & & &  &  \\
	 \hline
	\hline
 	&	\mbox{ $\Delta^{++} \rightarrow p \pi^+$}  &  	\mbox{ $\Delta^{+} \rightarrow n \pi^+$}  & 	\mbox{ $\Delta^{+} \rightarrow p \pi^0$}  &  	\mbox{ $\Delta^{0} \rightarrow n \pi^0$} & \mbox{ $\Delta^{+} \rightarrow n \pi^+$}  &  	\mbox{ $\Delta^{0} \rightarrow p \pi^-$}\\ 
 	\hline
	\mbox{DP} &	1 &  \frac{1}{3} &  \frac{2}{3}\cdot\sqrt{2}  &  \frac{2}{3}\cdot\sqrt{2} & - \frac{2}{3} &   \frac{2}{3} \\
 		 &	C_{i}^V & C_{i}^V & \frac{C_{i}^V}{2} (1-2 s_W^2 )   & \frac{C_{i}^V}{2} (1-2 s_W^2 )   & \frac{C_{i}^V}{2} (1-2 s_W^2 )   & \frac{C_{i}^V}{2} (1-2 s_W^2 ) \\
 		 &	C_{i}^A & C_{i}^A & \frac{C_{i}^A}{2} & \frac{C_{i}^A}{2} & \frac{C_{i}^A}{2} & \frac{C_{i}^A}{2}\\
	 \hline
	\mbox{CDP} & 1 &  3 & 2\cdot\sqrt{2} & 2\cdot\sqrt{2} &  2 & - 2\\
 		 &	C_{i}^V & C_{i}^V & \frac{C_{i}^V}{2} (1-2 s_W^2 )   & \frac{C_{i}^V}{2} (1-2 s_W^2 )   & \frac{C_{i}^V}{2} (1-2 s_W^2 )   & \frac{C_{i}^V}{2} (1-2 s_W^2 ) \\
 		 &	C_{i}^A & C_{i}^A & \frac{C_{i}^A}{2} & \frac{C_{i}^A}{2} & \frac{C_{i}^A}{2} & \frac{C_{i}^A}{2}\\ 
		  & & & & &  &  \\
 	\hline
	\end{array}
	\end{equation*}
	}
$s_W^2=\sin^2\theta_W$
\end{widetext}
The  $\mathcal{N}\to\Delta$ transition form factors ($C_i^{V,A}$) and $F_\rho$ are parameterized as in Ref.~\cite{Graczyk:2021oyl}.

\normalem
\bibliographystyle{apsrev4-1}
\bibliography{bib,bibdrat,bibdratbook,bibmoje}
\end{document}